\documentclass[
superscriptaddress,
amsmath,amssymb,
aps,
prl,
twocolumn
]{revtex4-1}

\usepackage{wrapfig}

\usepackage{tikz}
\usepackage{graphicx}
\graphicspath{{figures/}}
\usepackage{dcolumn}
\usepackage{bm}
\usepackage{hyperref}
\usepackage{multirow}
\usepackage{array}
\usepackage{booktabs}
\usepackage{ctable}
\usepackage{upgreek}
\usepackage{epsfig,psfrag,subfigure,amsopn}
\usepackage{mathrsfs}
\usepackage{amssymb}
\usepackage{amsbsy}
\usepackage{color}
\usepackage{cancel}

\usepackage{pifont}
\usepackage{marginnote}
\usepackage{float}

\definecolor{dgreen}{rgb}{0,0.7,0}

\newcommand{\titlename}{Decorrelation of a leader by the increasing number of followers}
\newcommand{\bea}{\begin{eqnarray}}
\newcommand{\eea}{\end{eqnarray}}

\begin{document}
	
	\title{\titlename}

%

\author{Satya N. Majumdar}
\address{LPTMS, CNRS, Univ. Paris-Sud, Universit\'e Paris-Saclay, 91405 Orsay, France}
\author{Gr\'egory Schehr}
\address{Sorbonne Universit\'e, Laboratoire de Physique Th\'eorique et Hautes Energies, CNRS UMR 7589, 4 Place Jussieu, 75252 Paris Cedex 05, France}	
	
\date{\today}
	
\begin{abstract}
We compute the connected two-time correlator of the maximum $M_N(t)$ of $N$ independent Gaussian stochastic processes (GSP) characterised by a 
common correlation coefficient $\rho$ that depends on the two times $t_1$ and $t_2$. We show analytically that this correlator, for fixed times $t_1$ and $t_2$, decays for large $N$ as a power law $N^{-\gamma}$ (with logarithmic corrections) with a decorrelation exponent $\gamma = (1-\rho)/(1+ \rho)$ that depends only on $\rho$, but otherwise is universal for any GSP. We study several examples of physical processes including the fractional Brownian motion (fBm) with Hurst exponent $H$ and the Ornstein-Uhlenbeck (OU) process. For the fBm, $\rho$ is only a function of $\tau = \sqrt{t_1/t_2}$ and we find an interesting ``freezing'' transition at a critical value $\tau= \tau_c=(3-\sqrt{5})/2$. For $\tau < \tau_c$, there is an optimal $H^*(\tau) > 0$ 
that maximises the exponent $\gamma$ and this maximal value freezes to $\gamma= 1/3$ for $\tau >\tau_c$. For the OU process, we show that $\gamma = {\rm tanh}(\mu \,|t_1-t_2|/2)$ where $\mu$ is the stiffness of the harmonic trap. Numerical simulations confirm our analytical predictions.  
\end{abstract}	
	
\maketitle

The time evolution of several complex systems in nature is often governed, not by the bulk
of the system, but rather by the outliers or extremes that live at the edges of the system. 
Examples include bacterial infections \cite{geisel}, evolution models \cite{krug_carl,krug_jain,jepsen}, self-propelled particles in active 
matter \cite{mori}, barrier crossing in chemical reactions \cite{kramers,hanggi}, search processes \cite{benichou,resetting}, spikes of correlation matrices of financial data \cite{bbp}, etc. The laws that govern the dynamics of such extremes are typically unknown and are very different from the laws
that govern the dynamics of the bulk. Consider for instance an assembly of $N$ particles in one-dimension with positions $\{X_i(t)\}$, $i=1, \cdots, N$. Let 
\bea \label{def_MN}
M_N(t) = \max\{X_1(t) ,X_2(t), \cdots ,X_N(t) \}
\eea
denote the position of the extreme or the leader at time $t$. The dynamics of $M_N(t)$ is in general very complicated, even when the particles are independent. This is because the label of the particle that takes the lead changes with time, introducing nontrivial correlations in the dynamics of $M_N(t)$. For independent particles, 
while at a fixed time $t$ the fluctuations of $M_N(t)$ can be described by the standard extreme value statistics (EVS) of independent variables, the evolution of $M_N(t)$ as a function of time is not described by the EVS. Characterizing the dynamics of $M_N(t)$, even for independent stochastic processes $\{X_i(t)\}$,  
is thus a challenging problem.  

This question of the extreme/leader dynamics appears quite naturally in many complex systems, such as growing networks \cite{Krap_Book,Krap_Redner2002}, $N$ independent particles on the line undergoing stochastic dynamics \cite{Krap_Book,Krap_Redner2002,B2010,BK2010,KMR2010,WMS2012,BK2014,KMS2014,BKL2015,BKR2016,Krap2021,OAM2022,Dean2023,pierre}, evolutionary dynamics in fitness landscapes \cite{krug_carl, krug_jain}, nonequilibrium dynamics of a freely expanding gas of hard-point particles \cite{jepsen}, etc. For example, in the case of a growing network, a leader at a given time is the node with the maximum degree of connections at that time. In these systems, questions such as how many lead changes occur in a given time or how long does a leader stay a leader before being taken over by a follower etc have been studied \cite{Krap_Book,Krap_Redner2002,B2010,BK2010,BKL2015,Krap2021}. In this paper, we address a different question. We consider $N$ independent stochastic processes $\{X_i(t)\}$, for $i=1, \dots, N$, and study a different observable associated to the maximum $M_N(t)$, namely, the connected correlator, $\langle M_N(t_1) M_N(t_2)\rangle - \langle M_N(t_1) \rangle \langle M_N(t_2) \rangle$, at two different times $t_1$ and $t_2$. Here our principal interest is to investigate how this correlator behaves as a function of increasing $N$ for fixed $t_1$ and $t_2$. As $N$ increases, one expects indeed that the particle that achieves the maximum at time $t_2 > t_1$ is typically different from the one at time $t_1$. Since the particles are independent, this correlator, for fixed $t_1$ and $t_2$, will thus decrease with increasing $N$. The question is: how does this ``decorrelation'' occur as a function of increasing $N$? Note that this question is different from the auto-correlation of the running maximum for one or more random walks studied in the literature \cite{BKMO2016,BKR2016}.

\begin{figure}[t]
\includegraphics[width = 0.9\linewidth]{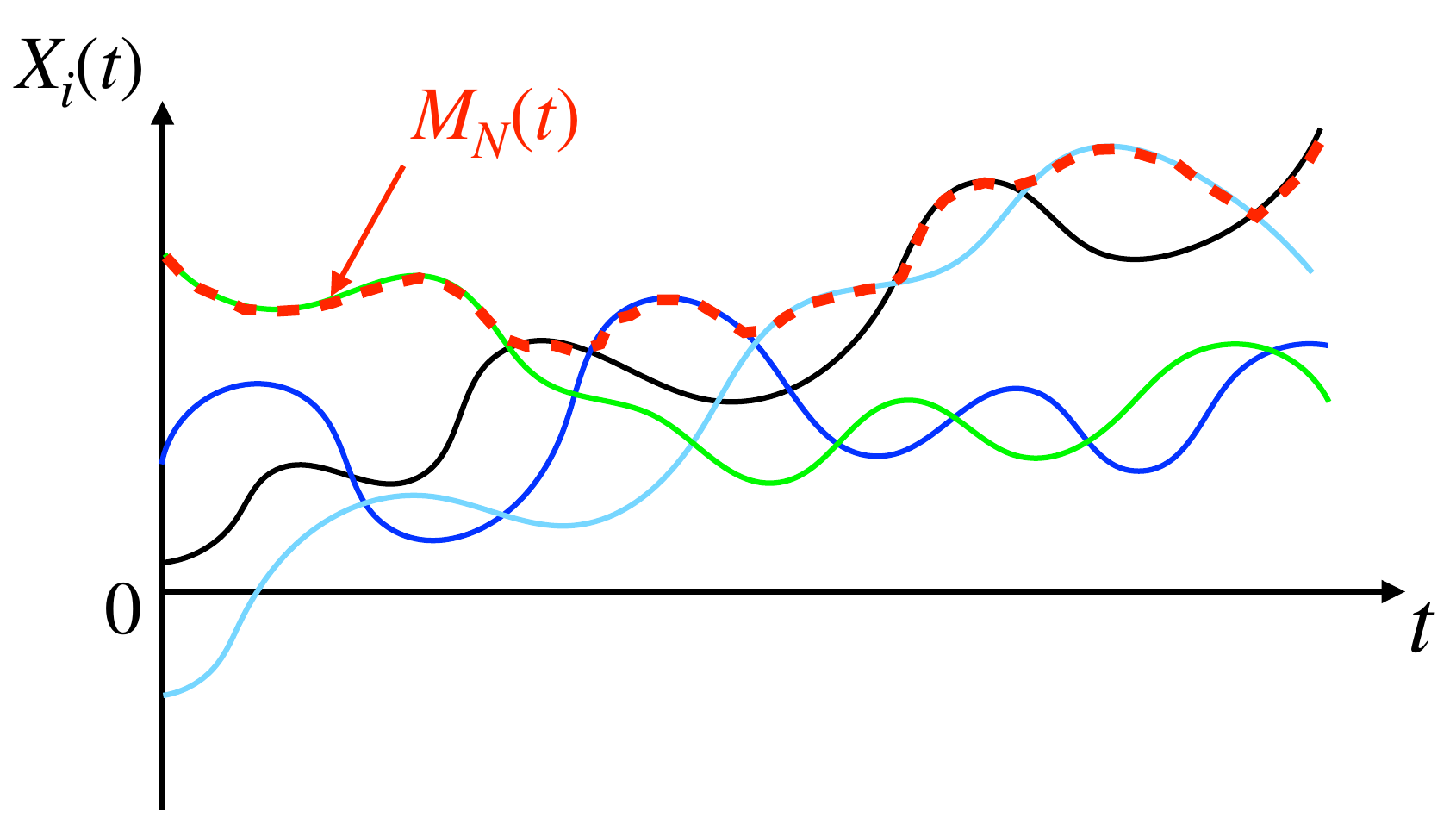}
\caption{A schematic figure of the trajectories of $N=4$ independent Gaussian stochastic processes with the same covariance function. The maximum (leader) $M_N(t)$ is the upper enveloppe of the four trajectories shown by the dashed red line.}\label{Fig_intro}
\end{figure}
In this paper, we consider $N$ independent Gaussian stochastic processes (GSPs) $\{X_i(t) \}$. Many processes in both natural (physical, chemical, biological) as well as artificial (finance, computer science) systems can be modelled by Gaussian processes~\cite{Chaturvedi,AdlerBook,MacKay,Rasmussen}. The prominent examples include Brownian motion, Ornstein-Uhlenbeck process, fractional Brownian motion etc. A GSP $X(t)$ is completely characterized by its covariance matrix, i.e., its joint distribution at $M$ different times $t_1, t_2, \ldots, t_M$ can be expressed as
\begin{equation}  \label{JPDF}
P[\{X(t_i)=x_i\}] = \frac{1}{2\pi \sqrt{{\rm det} \, C}}e^{-\frac{1}{2} {\sum_{i,j}} C^{-1}_{ij} \, x_i x_j}\;,
\end{equation}    
where $C_{ij} = \langle X(t_i) X(t_j)\rangle$ is the $M \times M$ covariance matrix and ${\rm det} C$ denotes its determinant. We now consider $N$ such independent GSP's, but all of them with the identical covariance function $C_{ij}$. We note that, since the $N$ particles are independent, the correlator $\langle M_N(t_1) M_N(t_2)\rangle - \langle M_N(t_1) \rangle \langle M_N(t_2) \rangle$ is nonzero if and only if the leader at time $t_1$ coincides with the leader at time $t_2$. In other words, 
\begin{equation} \label{relationS_N}
\langle M_N(t_1) M_N(t_2)\rangle - \langle M_N(t_1) \rangle \langle M_N(t_2) \rangle = C_{12}\; S_N(t_1, t_2) \;,
\end{equation}
where $S_N(t_1, t_2)$ is the probability that the leader at time $t_1$ is the same as the leader at time $t_2$. Consequently the scaled two-time correlator 
\begin{equation} \label{def_C}
\hspace*{0.2cm}{\cal C}_N(t_1,t_2) = \frac{\langle M_N(t_1) M_N(t_2)\rangle - \langle M_N(t_1) \rangle \langle M_N(t_2) \rangle}{2 \sqrt{C_{11} C_{22}}} 
\end{equation}
is given by
\bea \label{def_C2}
\hspace*{0.cm}{\cal C}_N(t_1,t_2)  = \frac{\rho}{2} \, S_N(t_1,t_2) \;,
\eea
where $0 \leq \rho \leq 1$ denotes the correlation coefficient
\bea \label{def_rho}
 \rho = \frac{C_{12}}{\sqrt{C_{11} C_{22}}} \;.
\eea
The quantity $S_N(t_1, t_2)$, or equivalently the scaled correlator ${\cal C}_N(t_1,t_2) $ in Eq. (\ref{def_C}), contains the essential informations about how the leader decorrelates, for fixed $t_1$ and $t_2$, as a function of increasing size $N$. The purpose of this Letter is precisely to compute analytically this scaled correlator for large $N$ and show it has a rather rich and interesting behavior as a function of increasing $N$.

{\it Summary of the main results.} Let us first summarize our main results for large $N$ that can be conveniently expressed in terms of 
\bea \label{N_tilde}
\tilde N = \frac{N}{2 \sqrt{\pi}} \;.
\eea
We show that, for large $N$ and $0 < \rho < 1$, the correlator decays as 
\bea \label{main_res}
{\cal C}_N(t_1,t_2) \underset{N \to \infty}{\approx} \frac{B(\rho)}{\left( \ln \tilde N\right)^{\frac{1+2\rho}{1+\rho}}}\, \tilde N^{-\gamma}  \;,
\eea
where the decorrelation exponent $\gamma$ and the amplitude $B(\rho)$ are given explicitly by 
\begin{equation} \label{gamma_rho}
\gamma = \frac{1-\rho}{1+\rho}  \quad, \quad B(\rho) = \frac{1}{8 \sqrt{\pi}} \frac{(1+\rho)^{3/2}}{\sqrt{1-\rho}}\, \Gamma^2\left( \frac{1}{1+\rho}\right) \;.
\end{equation}
The dependence of ${\cal C}_N(t_1,t_2)$ on $t_1$ and $t_2$ in Eq.~(\ref{main_res}) appears only through the correlation coefficient $\rho$ in (\ref{gamma_rho}).
Thus the decorrelation with increasing $N$ occurs rather slowly, with a leading power law decay characterized by the decorrelation exponent $\gamma$ that depends continuously on $\rho$, and with an additional logarithmic correction. Since $\rho$ can be explicitly computed for several GSPs (see later), one can then compute the exponent $\gamma$ explicitly. We also derive the limiting behavior of the correlator in the two extreme limits $\rho \to 0$ and $\rho \to 1$.

We will provide an outline of the derivation of our main result in Eq. (\ref{main_res}) at the end of the Letter (with more details in the Supplementary Material \cite{SM}). Here, we will first discuss the different physical implications and applications of the principal result in Eq. (\ref{main_res}) valid for $0<\rho<1$ 
and then discuss the two limiting cases $\rho \to 0$ and $\rho \to 1$.

\begin{figure}[t]
\includegraphics[width = \linewidth]{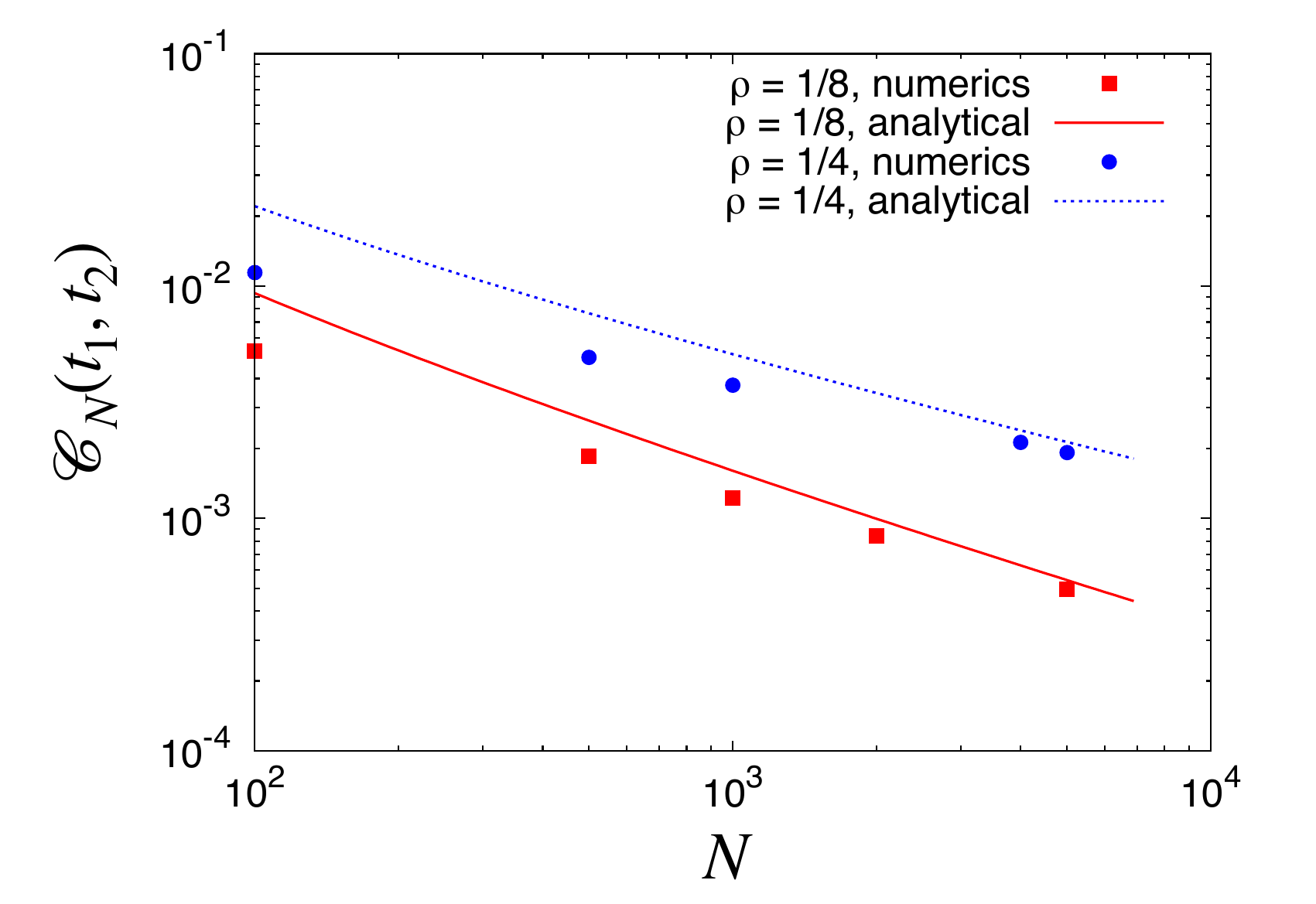}
\caption{Plot of ${\cal C}_N(t_1,t_2)$ for $N$ Brownian motions with $D=1$ as a function of $N$ for two different values of $\rho = 1/8$ (with $t_1=1, t_2 = 8$) and $\rho = 1/4$ (with $t_1=1, t_2=4$) in red and blue respectively. The lines correspond to the analytical prediction in Eqs. (\ref{main_res})-(\ref{gamma_rho}). The convergence to the asymptotic large $N$ formula is rather slow.}\label{Fig_num}
\end{figure}
{\it Examples.} We start with the simplest physical example of a Gaussian process, namely the Brownian motion. We consider  
a set of $N$ independent Brownian motions in one-dimension. A single Brownian motion is a GSP with covariance function $C_{12}= 2 D \min (t_1, t_2)$ where $D$ is the diffusion constant. Consequently $C_{11} = 2 D t_1$ and $C_{22} = 2 D t_2$. Hence the correlation coefficient $\rho$, for $t_1<t_2$, is given~by 
\bea \label{rho_BM}
\rho = \sqrt{\tau} \quad {\rm with} \quad \tau = t_1/t_2 \;. 
\eea   
Consequently, the decorrelation exponent $\gamma$ in Eq. (\ref{gamma_rho}) is given by 
\bea \label{gamma_BM}
\gamma = \frac{1-\sqrt{\tau}}{1+ \sqrt{\tau}} \;.
\eea
When $t_2 \gg t_1$, we see from Eq. (\ref{gamma_BM}) that $\rho \to 0$. To verify this prediction, we performed numerical simulations and found agreement with these results (see Fig. \ref{Fig_num}). 

Another well know GSP is the Ornstein-Uhlenbeck process, corresponding to an overdamped particle moving in a
harmonic potential with stiffness $\mu$. In the stationary state, the autocorrelator decays exponentially with the time difference as $\langle X(t_1) X(t_2) \rangle \sim e^{-\mu |t_1 - t_2|}$. Thus the correlation coefficient $\rho$ is given by $\rho = e^{-\mu |t_1-t_2|}$. Correspondingly, the decorrelation exponent $\gamma$ in Eq. (\ref{gamma_rho}) is given by
\bea \label{gamma_OU}
\gamma = \frac{1-\rho}{1+\rho} = {\rm tanh}\left( \frac{\mu\,|t_1 - t_2|}{2}\right) \;.
\eea
This analytical prediction may possibly be measured in optical traps experiments for a dilute gas of $N$ particles.   
In this context, our results predict that the scaled correlation function ${\cal C}_N(t_1,t_2)$ of the position of the rightmost particle 
in the trap, for fixed $t_1$ and $t_2$, will decay as a power law to leading order with a nontrivial exponent $\gamma$ given in Eq. (\ref{gamma_OU}).

\begin{figure}[t]
\includegraphics[width = \linewidth]{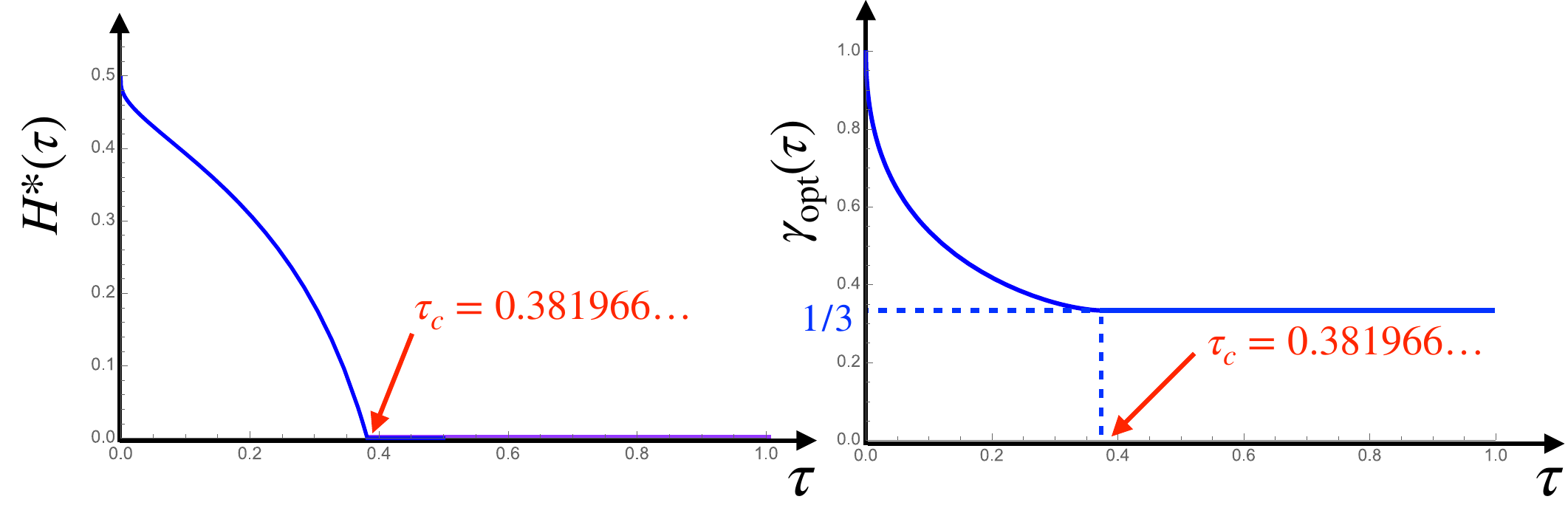}
\caption{{\bf Left:} The Hurst exponent $H^{*}(\tau)$ of the fBm, that maximises the decorrelation exponent $\gamma$, plotted as a function of $\tau$. It vanishes linearly  at $\tau_c =  \frac{3-\sqrt{5}}{2} = 0.381966\ldots$. {\bf Right:} the optimised exponent $\gamma_{\rm opt}(\tau)$ vs $\tau$. It approaches the value $1/3$ quadratically as $\tau \to \tau_c$, as in Eq. (\ref{gam_opt_quad}).}\label{Fig_Hstar}
\end{figure}

Yet another GSP that has been widely studied with many applications ranging from fractals \cite{Mandelbrot,fractals}, climate dynamics \cite{climate}, 
anomalous diffusion \cite{Weiss,Klafter} to finance \cite{Oksendal} is the celebrated fractional Brownian motion (fBm) \cite{Mandelbrot,Kolmo,Beran,Perrin}. An fBm is a GSP with correlator 
\bea \label{correlfBM}
C_{12} = D \left(t_1^{2H} + t_2^{2H}- |t_1-t_2|^{2H}  \right) \;,
\eea 
where $0<H<1$ is the Hurst exponent characterizing the fBm. The value $H=1/2$ corresponds to the standard 
Brownian motion that separates the subdiffusive ($0<H<1/2$) and the superdiffusive regime ($1/2<H<1$). A well known example of a sub-diffusive fBm with Hurst exponent $H=1/4$ corresponds to the position of a tagged monomer in a one-dimensional Rouse polymer chain \cite{Panja,Taloni} or the height of an Edwards-Wilkinson interface at a fixed position in space \cite{Krug_FBM,Bray}. In the latter context, one can create a family of interfaces whose dynamics is governed by fBm with a tunable Hurst exponent $H$ by varying the dynamical exponent and the dimension $d$ of the interface \cite{Krug_FBM}. 
For the fBm with Hurst exponent $H$, using Eq. (\ref{correlfBM}), the correlation coefficient $\rho$ is given by 
\bea \label{rho_tau}
\rho = \frac{1 + \tau^{2H}-(1-\tau)^{2H}}{2 \tau^H} \;,
\eea 
where $\tau = \sqrt{t_1/t_2}$ for $t_1 \leq t_2$. The decorrelation exponent $\gamma = (1-\rho)/(1+\rho)$ from Eq. (\ref{gamma_rho}) thus depends on two parameters $0<\tau<1$ 
and $0 < H< 1$. As a function of these two parameters, the exponent $\gamma(\tau, H)$ undergoes 
a very interesting and unexpected "phase transition". Indeed, by analysing the explicit formula of $\gamma$, we find that there is a critical value 
\bea \label{tau_c}
\tau_c = \frac{3-\sqrt{5}}{2} = 0.381966\ldots \;, 
\eea
such that for $\tau > \tau_c$, the decorrelation exponent $\gamma$ decreases monotonically with increasing $H$, starting from $\gamma(\tau,H=0) = 1/3$. Thus in this regime, the most sub-diffusive fBm (with $H=0$) maximises the decorrelation exponent $\gamma$, with value $1/3$. However, for $\tau < \tau_c$, the exponent $\gamma$ is maximal at a nonzero value $H = H^*(\tau)$. This optimal Hurst exponent $H^*(\tau)$, as a function of $\tau$, decreases from its value $1/2$ at $\tau = 0$ and vanishes at $\tau_c$ linearly as $H^*(\tau) \approx a (\tau_c - \tau)$ where the prefactor $a$ has a nontrivial value
\bea \label{a}
a = \frac{\sqrt{5}(3 + \sqrt{5})}{[4 {\rm Arcoth}(\sqrt{5})]^2} = 3.16007 \ldots  
\eea
Therefore the optimal decorrelation exponent $\gamma_{\rm opt}(\tau) = \gamma(\tau,H^*(\tau))$ decreases as a function of increasing $\tau$, starting from $\gamma_{\rm opt}(\tau=0) = 1$ and freezes to the value $\gamma_{\rm opt}(\tau) = 1/3$ for $\tau \geq \tau_c$ where $\tau_c$ is given in (\ref{tau_c}). Close to $\tau = \tau_c$, the optimal exponent behaves as 
\bea \label{gam_opt_quad}
\gamma_{\rm opt}(\tau) \approx \frac{1}{3} + b (\tau_c-\tau)^2 \;,
\eea
where $b$ has an explicit but complicated expression, with value $b = 4.11098 \ldots $. In Fig. \ref{Fig_Hstar} we plot both $H^*(\tau)$ and $\gamma_{\rm opt}(\tau)$ as a function of $\tau$.

{\it Two limiting cases $\rho \to 0$ and $\rho \to 1$.} We next consider the two extreme limits $\rho \to 0$ and $\rho \to 1$. The former case corresponds to two times $t_1$ and $t_2$ that are widely separated, while the latter case corresponds to closely separated times $t_1$ and $t_2$. In the limit $\rho \to 0$, we find that the leading term in Eq. (\ref{main_res}) actually behaves as $1/N$ for large $N$. More precisely, we find \cite{SM}
\bea \label{rho_0}
\hspace*{-0.cm}{\cal C}_N(t_1,t_2) \underset{N \to \infty}{\simeq} \frac{\rho}{2\,N} \;.
\eea
Using this result in Eq. (\ref{def_C2}) predicts that $S_N(t_1,t_2) \approx 1/N$. This is indeed what one would expect since, in this limit $\rho \to 0$, the set of positions $\{X_1(t_1), \cdots, X_N(t_1)\}$  at time $t_1$ are almost uncorrelated from those at time $t_2$ and hence the probability that the leader at $t_1$ is also the leader at $t_2$ is exactly $1/N$.

In the opposite limit when $t_2 \to t_1$, the correlation coefficient $\rho \to 1$. In this limit, the large $N$ correlator ${\cal C}_N(t_1, t_2)$ in Eq. (\ref{main_res}) diverges, but in a universal way for any GSP
\begin{equation} \label{rho_1}
{\cal C}_N(t_1,t_2) \underset{N \to \infty}{\sim}  \sqrt{\frac{\pi}{8 (1- \rho)}} \, \left(\ln{\tilde N}\right)^{-3/2} \quad, \quad t_2 \to t_1 \;.
\end{equation} 
This divergence can be understood physically as follows. When $t_2 \to t_1$, it is clear that the correlator ${\cal C}_{N}(t_1,t_2)$ approaches the scaled variance of the maximum at fixed time $t_1$. It is well known from the theory of EVS for Gaussian random variables \cite{Gumbel_book,EVS_review} that, for large $N$, the scaled maximum $M_N(t)$ at fixed time $t$ behaves for large $N$ as (see also \cite{SM})
\bea \label{Gumbel}
\tilde M_1 = \frac{M_N(t_1)}{\sqrt{2 C_{11}}} \underset{N \to \infty}{\longrightarrow} a_N + \frac{1}{2 a_N}\, Z_1 \;,
\eea
where $Z_1$ is an $N$-independent random variable distributed via the Gumbel law
\bea \label{Gumbel_law}
{\rm Prob}(Z_1 \leq z) = e^{-e^{-z}} \;,
\eea
and the scale factor $a_N$ satisfies \cite{EVS_review}
\bea \label{def_aN}
e^{-a_N^2} = \frac{2 \sqrt{\pi}}{\tilde N}\, a_N \;,
\eea
which, for large $N$, behaves as
\bea \label{aN_largeN}
a_N \underset{N \to \infty}{\approx} \sqrt{\ln \tilde N} - \frac{\ln (\ln \tilde N)}{4 \sqrt{\ln \tilde N}} \;.
\eea
Consequently the scaled variance ${\rm Var}(\tilde M_1) = \langle \tilde M_1^2\rangle - \langle \tilde M_1\rangle^2$, from Eq. (\ref{Gumbel}), is expected to behave, for large $N$, as~\cite{SM}
\bea \label{var_Gumbel}
{\rm Var}(\tilde M_1) \approx \frac{\pi^2}{24\,a_N^2} \approx \frac{\pi^2}{24\, \ln \tilde N} \;.
\eea
Comparing this to Eq. (\ref{rho_1}) in the limit $t_1 \to t_2$, one finds that the two regimes match with each other when $1 - \rho \sim O(1/a_N^2) \sim O(1/\ln \tilde N)$. This suggests that if we set 
\bea \label{scaling_rho1}
\rho = 1 - \frac{\nu}{a_N^2}  \;,
\eea
where $\nu = O(1)$ when $N \to \infty$, then the limiting process of the scaled maximum should be independent of $N$ and just parametrised by $\nu$. This limiting process has been studied both in the mathematics \cite{kabluchko,BR1977,HR1989} and physics \cite{pierre} literature,  
but the result for finite $1-\rho$ in (\ref{rho_1}) was not known. In fact, the limit $\nu \to \infty$ should correspond to Eq. (\ref{rho_1}), while the opposite limit $\nu \to 0$ should correspond to (\ref{var_Gumbel}). Indeed we show in \cite{SM} that the joint cumulative distribution of the scaled maximum, with the correlation coefficient scaling as in (\ref{scaling_rho1}), approaches a limiting $N$-independent form for large $N$ parametrised by $\nu$. This limiting form interpolates smoothly between the two limits respectively in Eqs. (\ref{rho_1}) and (\ref{var_Gumbel}).

{\it Sketch of the derivation}. We now briefly outline the steps leading to our main results in Eq. (\ref{main_res}) (more details are provided in \cite{SM}). We start with a single GSP with covariance function $C_{ij}$. For fixed $t_1, t_2$ and setting $X(t_1)=x_1$ and $X(t_2)=x_2$, the joint distribution of $x_1$ and $x_2$, from Eq. (\ref{JPDF}) with $M=2$, can be expressed as
\bea \label{JPDF_M2}
\hspace*{-0.3cm} P(x_1, x_2) =A_1\,e^{-\frac{x_1^2}{2(1-\rho^2)C_{11}}-\frac{x_2^2}{2(1-\rho^2)C_{22}}+ \frac{\rho}{1-\rho^2} \frac{x_1 \,x_2}{\sqrt{C_{11} C_{22}}}} 
\eea
where $A_1 =  \frac{1}{2 \pi \sqrt{C_{11}\,C_{22}(1-\rho^2)}}$ and $\rho$ is defined in Eq.~(\ref{def_rho}). Next we consider $N$ independent GSPs $\{X_i(t)\}$ for $i=1, \cdots, N$ with identical covariance function $C_{ij}$ and denote their maximum by $M_N(t)$ as in Eq. (\ref{def_MN}). A convenient starting point for computing the correlator ${\cal C}_N(t_1,t_2)$ in Eq. (\ref{def_C}) is to study the joint cumulative distribution of the maximum 
\bea \label{def_FN}
\hspace*{-0.6cm}F_N(M_1, M_2) = {\rm Prob.}\left[M_N(t_1) \leq M_1, M_N(t_2) \leq M_2  \right].
\eea 
This event requires that each of the $N$ processes at $t_1$ and $t_2$ have values smaller than $M_1$ and $M_2$ respectively. Using 
 the independence of the GSPs, we can then write 
\begin{equation} \label{FN_2}
F_N(M_1, M_2) = \left[\int_{-\infty}^{M_1} dx_1 \,  \int_{-\infty}^{M_2} dx_2 \, P(x_1, x_2) \right]^N \;.
\end{equation}
Injecting the expression for $P(x_1, x_2)$ from (\ref{JPDF_M2}), and rearranging, we find that $F_N(M_1, M_2)$ can be expressed as
\begin{equation}  \label{scaled_FN}
F_N(M_1, M_2) = {\cal F}_N \left(m_1 =\frac{M_1}{\sqrt{2 C_{11}}}, m_2 =\frac{M_2}{\sqrt{2 C_{22}}}, \rho  \right)
\end{equation}
where  
\begin{equation} \label{calF_N}
{\cal F}_N(m_1, m_2, \rho) = \left(1 - \frac{1}{2} {\rm erfc}(m_1) - \frac{1}{2} {\rm erfc}(m_2) + I_N\right)^N
\end{equation}
with
\bea \label{def_I}
I_N = \int_{m_1}^\infty \frac{dy_1}{2\sqrt{\pi}} e^{-y_1^2} \, {\rm erfc}\left(\frac{m_2 - \rho\, y_1}{\sqrt{1- \rho^2}} \right) \;.
\eea
The next step is to substitute for the values of the scaled maxima $m_{1,2} = a_N + 1/(2 a_N)\, z_{1,2}$ in Eqs. (\ref{calF_N}) and (\ref{def_I}), where $z_{1,2}$ are $O(1)$ and $a_N$ is given in Eqs. (\ref{def_aN}) and (\ref{aN_largeN}) for large $N$.  Then we perform the asymptotic analysis of Eq. (\ref{calF_N}) for large $N$ (see \cite{SM} for details). From this cumulative distribution for large $N$ we then extract the connected correlation function $\langle Z_1 Z_2 \rangle - \langle Z_1 \rangle \langle Z_2 \rangle$. Finally the scaled correlator ${\cal C}_N(t_1,t_2) = \langle \tilde M_1 \tilde M_2 \rangle - \langle \tilde M_1 \rangle \langle \tilde M_2 \rangle$ is given by
\bea \label{C_z1z2}
{\cal C}_N(t_1,t_2) \approx \frac{1}{4 a_N^2} \left[\langle Z_1 Z_2 \rangle - \langle Z_1 \rangle \langle Z_2 \rangle \right] \;,
\eea 
where we used the relations $\tilde M_{1,2} = a_N + 1/(2 a_N)\, Z_{1,2}$. Finally, using the asymptotic behavior of $a_N$ in Eq.  (\ref{aN_largeN}) leads to our result in Eqs. (\ref{main_res}) and (\ref{gamma_rho}). 

{\it Conclusion.} In this Letter, we have shown that the scaled connected two-time correlator of the maximum $M_N(t)$ of $N$ independent GSPs with a common correlation coefficient $\rho$, for fixed times $t_1$ and $t_2$, decays with increasing $N$ as a power law $N^{-\gamma}$ (with logarithmic corrections) with a decorrelation exponent $\gamma = (1-\rho)/(1+ \rho)$ that depends only on $\rho$, but otherwise is universal for any GSP. This analytical prediction is verified in numerical simulations. We have shown that this decorrelation exponent $\gamma$ has nontrivial behaviors in several physical stochastic processes, such as the fractional Brownian motion (fBm) with Hurst exponent $0<H<1$ and the Ornstein-Uhlenbeck (OU) process. For the fBm, it turns out that there is an optimal Hurst exponent $H^*$ that maximises the exponent $\gamma$ and $H^*$ vanishes for $\tau = \sqrt{t_1/t_2} > \tau_c=(3-\sqrt{5})/2$. The OU process corresponds to a particle diffusing in a harmonic trap with stiffness $\mu$ and in this case, the decorrelation exponent $\gamma = {\rm tanh}(\mu \,|t_1-t_2|/2)$ may possibly be measurable in optical trap experiments. This work opens up several future directions, including the calculation of this new decorrelation exponent $\gamma$ for non-Gaussian processes, as well as for interacting particles.

{\it Acknowledgments.} We thank D. S. Dean, P. Le Doussal, B. Meerson and O. Vilk for useful discussions. 
We acknowledge support from ANR, Grant No. ANR- 23-CE30-0020-01 EDIPS.

\end{document}


\title{\titlename}

%

\author{Satya N. Majumdar}
\address{LPTMS, CNRS, Univ. Paris-Sud, Universit\'e Paris-Saclay, 91405 Orsay, France}
\author{Gr\'egory Schehr}
\address{Sorbonne Universit\'e, Laboratoire de Physique Th\'eorique et Hautes Energies, CNRS UMR 7589, 4 Place Jussieu, 75252 Paris Cedex 05, France}	
	
\date{\today}
	
\begin{abstract}
We provide some details of the derivation of the results presented in the main text. 
\end{abstract}	
	
\maketitle

\begin{figure}[t]
\includegraphics[width = 0.5\linewidth]{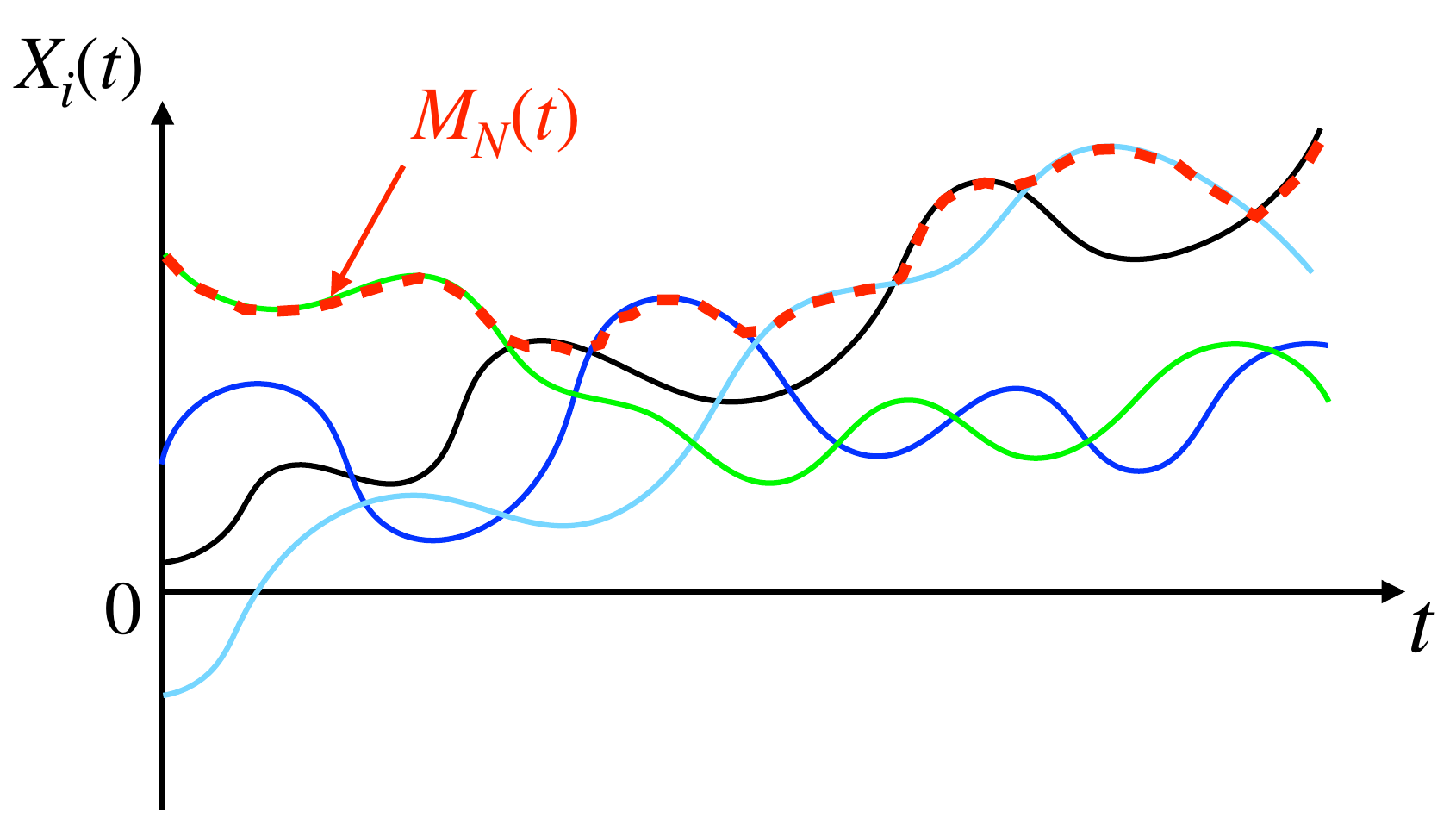}
\caption{A schematic figure of the trajectories of $N=4$ independent Gaussian stochastic processes with the same covariance function. The maximum (leader) $M_N(t)$ is the upper enveloppe of the four trajectories shown by the dashed red line.}\label{Fig_intro}
\end{figure}

\section{Basic formalism for Gaussian stochastic processes}\label{sec:gen}

A Gaussian stochastic process (GSP) is completely characterized by its covariance matrix, i.e., its joint distribution at $M$ different times $t_1, t_2, \ldots, t_M$ can be expressed as \cite{Chaturvedi}
\begin{equation}  \label{JPDF}
P[\{X(t_1)=x_1 , X(t_2) = x_2, \cdots, X(t_M) = x_M\}] = \frac{1}{2\pi \sqrt{{\rm det} \, C}}e^{-\frac{1}{2} {\sum_{i,j}} C^{-1}_{ij} \, x_i x_j}\;,
\end{equation}    
where $C_{ij} = \langle X(t_i) X(t_j)\rangle$ is the $M \times M$ symmetric covariance matrix and ${\rm det} \, C$ denotes its determinant. For example, with $M=1$, one obtains the one-point marginal distribution as 
\bea \label{marg}
P(X(t_k) = x_k) = \frac{1}{\sqrt{2 \pi C_{kk}}} e^{- \frac{x_k^2}{2\,C_{kk}}} \;,
\eea
where $C_{kk}$ is just the variance at time $t_k$. Similarly, for the two-point function, we need to consider the $2 \times 2$ covariance matrix
\bea \label{cov_mat}
C = \begin{bmatrix}
C_{11} & C_{12} \\
C_{12} & C_{22}
\end{bmatrix}
\eea 
whose determinant is simply ${\rm det C} = C_{11} C_{22} - C_{12}^2$. Consequently, the inverse of the covariance matrix is also $2 \times 2$ and is given by
\bea \label{C_inv}
C^{-1} = \frac{1}{\left(C_{11} C_{22} - C_{12}^2\right)}
\begin{bmatrix}
C_{22} & -C_{12} \\
-C_{12} & C_{11}
\end{bmatrix} \;.
\eea 
Substituting this in Eq. (\ref{JPDF}) gives the two-point joint distribution of the GSP
\bea \label{JPDF_2}
P(x_1, x_2) = \frac{1}{2\pi\sqrt{C_{11} C_{22} - C_{12}^2}} \,\exp{\left(-\frac{1}{2} \frac{C_{22}}{C_{11} C_{22} - C_{12}^2} x_1^2 - \frac{1}{2} \frac{C_{11}}{C_{11} C_{22} - C_{12}^2}x_2^2 + \frac{C_{12}}{C_{11} C_{22} - C_{12}^2} x_1\, x_2 \right)} \;.
\eea
In terms of the correlation coefficient 
\bea \label{def_rho}
\rho = \frac{C_{12}}{\sqrt{C_{11} C_{22}}}
\eea 
the two-point joint distribution in Eq. (\ref{JPDF_2}) can be conveniently expressed as
\bea \label{JPDF_M3}
\hspace*{-0.3cm} P(x_1, x_2) = \frac{1}{2 \pi \sqrt{C_{11}\,C_{22}(1-\rho^2)}}\,\exp{\left(-\frac{x_1^2}{2(1-\rho^2)C_{11}}-\frac{x_2^2}{2(1-\rho^2)C_{22}}+ \frac{\rho}{1-\rho^2} \frac{x_1 \,x_2}{\sqrt{C_{11} C_{22}}}\right)} \;.
\eea
This is the result that was used in Eq. (26) of the main text. 

We now consider $N$ independent GSPs $\{X_1(t), X_2(t), \cdots, X_N(t) \}$, all sharing the same covariance matrix $C_{ij}$ and consider their maximum 
\bea \label{def_MN}
M_N(t) = \max\{X_1(t) ,X_2(t), \cdots, X_N(t) \} \;.
\eea
Since $\max$ is a nonlinear function, the extremal process $M_N(t)$ is clearly non-Gaussian. In the following two sub-sections, we derive the one-point and the two-point joint distribution functions of this extremal process $M_N(t)$.  

\subsection{One-point function}

We start with the one-point distribution of the maximum $M_N(t)$. It is convenient to consider the cumulative one-point distribution defined as
\bea \label{CDF_N1point}
{\rm Prob.}(M_N(t_k) \leq M_k ) = {\rm Prob.}(X_1(t_k)\leq M_k, X_2(t_k) \leq M_k \cdots, X_N(t_k)\leq M_k) \;.
\eea
Using the independence of the $X_i$'s and the one-point distribution (\ref{marg}) it follows that
\bea \label{CDF_N1point2}
{\rm Prob.}(M_N(t_k) \leq M_k )  = \left( \int_{-\infty}^{M_k} \frac{dx_k}{\sqrt{2 \pi C_{kk}}} \,e^{- \frac{x_k^2}{2\,C_{kk}}}   \right)^N \;.
\eea
It is then natural to introduce the random variable corresponding to the scaled maximum 
\bea \label{Mtilde_k}
\tilde M_k = \frac{M_N(t_k)}{\sqrt{2 C_{kk}}} \;, 
\eea
in terms of which Eq. (\ref{CDF_N1point2}) reads
\bea \label{CDF_N1point3}
{\rm Prob.}\left(\tilde M_k \leq m_k\right) = \left(1- \frac{1}{2} {\rm erfc}(m_k) \right)^N \;,
\eea
where $ {\rm erfc}(y) = (2/\sqrt{\pi})\int_{y}^\infty e^{-u^2}\,du$. It is useful to then recall the limiting behavior of the scaled maximum for large $N$. The theory of Extreme Value Statistics (EVS) asserts that, for large $N$, the scaled maximum $\tilde M_k$ has a deterministic and a noisy part that can be expressed as
\bea \label{EVS_largeN}
\tilde M_k \approx a_N + b_N \, Z_k \;,
\eea
where $a_N$ and $b_N$ are $N$-dependent scale factors, while $Z_k$ is a random variable of order $O(1)$ whose distribution is independent of $N$ for large $N$. We now substitute (\ref{EVS_largeN}) in Eq. (\ref{CDF_N1point3}) and use the asymptotic large argument behavior of the complementary error function
\bea \label{large_erfc}
{\rm erfc}(y) \approx \frac{1}{y \sqrt{\pi}} e^{-y^2} \;.
\eea
Substituting this asymptotic form in Eq. (\ref{CDF_N1point3}), one gets to leading order for large $N$
\bea \label{asympt_Mk}
{\rm Prob.}\left(\tilde M_k \leq m_k\right) = \left(1- \frac{1}{2} {\rm erfc}(m_k) \right)^N \approx e^{-\frac{N}{2\sqrt{\pi} m_k}e^{-m_k^2}} \;.
\eea
We then substitute 
\bea \label{mk_zk}
m_k = a_N + b_N\, z_k \;,
\eea
where $z_k$ is the value of the random variable $Z_k$ defined in Eq. (\ref{EVS_largeN}). One then finds that, in order that the right hand side (RHS) of (\ref{CDF_N1point3}) becomes independent of $N$ to leading order, one must have
\bea \label{aNbN}
b_N = \frac{1}{2 a_N} \quad {\rm and} \quad e^{-a_N^2} = \frac{2 \sqrt{\pi}}{N}\,a_N \;,
\eea
and the expression in Eq. (\ref{asympt_Mk}) becomes
\bea \label{rel_zk} 
{\rm Prob.}\left(\tilde M_k \leq m_k\right)  \approx e^{-e^{-z_k}}
\eea
where we used 
\bea \label{rel_zk2}
\frac{N}{2 \sqrt{\pi}\,m_k} \, e^{-m_k^2} \approx e^{-z_k} \;.
\eea
Solving for $a_N$ from the second relation in (\ref{aNbN}) shows that, for large $N$, 
\bea \label{aN_largeN}
a_N \underset{N \to \infty}{\approx} \sqrt{\ln \tilde N} - \frac{\ln (\ln \tilde N)}{4 \sqrt{\ln \tilde N}} \;,
\eea
where we have defined, for convenience, 
\bea \label{Ntilde}
\tilde N = \frac{N}{2 \sqrt{\pi}} \;.
\eea
With this appropriately chosen scale factors Eq. (\ref{CDF_N1point3}) reduces to an $N$-independent form for large $N$
\bea \label{Gumbel}
{\rm Prob.}(Z_k \leq z_k) = e^{-e^{-z_k}} \;.
\eea
This is known as the Gumbel cumulative distribution in the standard theory of EVS~\cite{Gumbel_book,EVS_review}. Consequently the probability distribution function (PDF) of $Z_k$ is given by
\bea \label{PDF_Gumbel}
{\rm Prob.}(Z_k = z_k) = e^{-z_k - e^{-z_k}} \;.
\eea
The corresponding mean and the variance are given by
\bea \label{mean_var_gumb}
\langle Z_k \rangle = \gamma_E \quad {\rm and} \quad {\rm Var}(Z_k) = \langle Z_k^2 \rangle - \langle Z_k \rangle^2 = \frac{\pi^2}{6} \;,
\eea
where $\gamma_E$ is the Euler constant. Subsequently, using Eqs. (\ref{EVS_largeN}) and (\ref{aN_largeN}), the mean and the variance of the scaled maximum are given by
\bea \label{mean_var_MN}
\left\langle \frac{M_N(t_k)}{\sqrt{2 C_{kk}}}  \right\rangle \approx a_N + \frac{\gamma_E}{2\,a_N} \approx \sqrt{\ln \tilde N} \quad {\rm and} \quad {\rm Var}\left(\frac{M_N(t_k)}{\sqrt{2 C_{kk}}} \right) \approx \frac{\pi^2}{24 \,a_N^2} \approx \frac{\pi^2}{24\, \ln(\tilde N)} \;,
\eea
where we used the leading term in the large $N$ behavior of $a_N$ in Eq. (\ref{aN_largeN}). This then provides a derivation of the results given in Eqs. (20)-(24) in the main text.

\subsection{Two-point joint distribution}

In this subsection, we want to investigate the two-point joint distribution of the extremal process $M_N(t)$. As in the case of the one-point function, it is convenient to study the joint cumulative distribution of the maximum at two times $t_1$ and $t_2$, defined as   
\bea \label{def_FN_SM}
\hspace*{-0.4cm}F_N(M_1, M_2) = {\rm Prob.}\left[M_N(t_1) \leq M_1, M_N(t_2) \leq M_2  \right] \,.
\eea 
This event requires that each of the $N$ processes at $t_1$ and $t_2$ have values smaller than $M_1$ and $M_2$ respectively. Using 
 the independence of the GSPs, we can then write 
\begin{equation} \label{FN_2}
F_N(M_1, M_2) = \left[\int_{-\infty}^{M_1} dx_1 \,  \int_{-\infty}^{M_2} dx_2 \, P(x_1, x_2) \right]^N \;.
\end{equation}
Injecting the expression for $P(x_1, x_2)$ from (\ref{JPDF_M3}), and rearranging, we find that $F_N(M_1, M_2)$ can be expressed as
\begin{equation}  \label{scaled_FN}
F_N(M_1, M_2) = {\cal F}_N \left(m_1 =\frac{M_1}{\sqrt{2 C_{11}}}, m_2 =\frac{M_2}{\sqrt{2 C_{22}}}, \rho  \right)
\end{equation}
where  ${\cal F}_N(m_1, m_2, \rho)$ is given by
\bea \label{FN_1}
{\cal F}_N(m_1, m_2, \rho) = \left[\frac{1}{\pi\sqrt{1-\rho^2}}\int_{-\infty}^{m_1} du_1 \int_{-\infty}^{m_2} du_2 \exp{\left(-\frac{1}{1-\rho^2}(u_1^2+u_2^2) + \frac{2\rho}{1-\rho^2} \,u_1\,u_2 \right)}  \right]^N \;.
\eea
For convenience, we now change the limits of integrations $\int_{-\infty}^{m_1} du_1 \to \int_{-\infty}^{\infty}du_1 - \int_{m_1}^{\infty}du_1$ and similarly for $u_2$. This gives
\begin{equation} \label{calF_N_SM}
{\cal F}_N(m_1, m_2, \rho) = \left(1 - \frac{1}{2} {\rm erfc}(m_1) - \frac{1}{2} {\rm erfc}(m_2) + I_N\right)^N
\end{equation}
where
\bea \label{I_SM1}
I_N = \frac{1}{\pi\,\sqrt{1-\rho^2}} \int_{m_1}^\infty du_1 \int_{m_2}^\infty du_2 \exp{\left( -\frac{1}{1-\rho^2}(u_1^2+u_2^2) + \frac{2\rho}{1-\rho^2} \,u_1\,u_2  \right)} \;.
\eea
Performing the Gaussian integral over $u_2$ explicitly, one gets a single integral 
\bea \label{def_I}
I_N = \int_{m_1}^\infty \frac{dy_1}{2\sqrt{\pi}} e^{-y_1^2} \, {\rm erfc}\left(\frac{m_2 - \rho\, y_1}{\sqrt{1- \rho^2}} \right) \;.
\eea
Note that the $N$-dependence of $I_N$ is only through $m_1$ and $m_2$ which do depend on $N$. This provides a derivation of Eqs. (30) and (31) in the main text. 

The next step is to provide the large $N$ asymptotic behavior of this single integral $I_N$. This requires a series of tricks. Hence we present it in a separate section. Once we have the large $N$ asymptotics of $I_N$, we can then analyse the cumulative joint distribution ${\cal F}_N(m_1, m_2, \rho)$ in (\ref{calF_N_SM}) for large $N$. Our goal is to derive a limiting scaling form for the two-point function ${\cal F}_N(m_1, m_2, \rho)$ for large $N$, as we did for the single point function in Eq. (\ref{Gumbel}). Finally, we will use this limiting scaling form to compute the two-point scaled correlator 
\begin{equation} \label{def_C_SM}
{\cal C}_N(t_1,t_2) = \frac{\langle M_N(t_1) M_N(t_2)\rangle - \langle M_N(t_1) \rangle \langle M_N(t_2) \rangle}{2 \sqrt{C_{11} C_{22}}} \;.
\end{equation}

\section{Large $N$ asymptotic analysis of $I_N$ in Eq. (\ref{def_I})}

In this section, we provide the large $N$ asymptotics of $I_N$ in Eq. (\ref{def_I}). This large $N$ analysis is not so easy since the $N$-dependence appears through $m_1$ and $m_2$. Note that $m_1$ appears in the lower limit of the integrand in Eq.~(\ref{def_I}), while $m_2$ appears inside the integrand -- this makes the large $N$ analysis a bit more tricky. To proceed, we first make a change of variable $y_1 = m_1 + u/(2 m_1)$. This gives
\bea \label{IN_large1}
N\, I_N = \frac{N}{2} \frac{e^{-m_1^2}}{2 \sqrt{\pi}\,m_1} \int_0^\infty du \, e^{-u - \frac{u^2}{4\,m_1^2}} \,{\rm erfc}\left( \frac{m_2 - \rho\,m_1}{\sqrt{1-\rho^2}} - \frac{\rho\,u}{2\,m_1 \sqrt{1-\rho^2}}\right) \;.
\eea 
Now using Eq. (\ref{rel_zk2}) for $m_1$, we can replace the prefactor outside the integral by $e^{-z_1}/2$. Furthermore, in the large $m_1$ limit, we ignore the term $u^2/(4m_1)$ in the argument of the exponential in (\ref{IN_large1}) as it leads to subleading corrections. This then leads to a simplified expression
 \bea \label{IN_large2}
N\, I_N \approx \frac{1}{2} e^{-z_1} \int_0^\infty du\, e^{-u} {\rm erfc} \left(\frac{m_2 - \rho\,m_1}{\sqrt{1-\rho^2}} - \frac{\rho \,u}{2 m_1 \sqrt{1-\rho^2}}\right) \;. 
\eea

Thus we managed to put both $m_1$ and $m_2$ inside the integrand and the limits of the integral in (\ref{IN_large2}) do not involve $m_1$ any more. But we are not done yet, because analysing the large $N$ behaviour of the integrand and then performing the integral is still not easy. Fortunately, this integral in Eq. (\ref{IN_large2}) can be explicitly performed using the following identity (obtained using Mathematica)
\bea \label{identity}
\int_0^\infty du\, e^{-u} \, {\rm erfc}\left(A - \frac{u}{B} \right) = e^{\frac{B^2}{4}-A\,B} {\rm erfc}\left(\frac{B}{2}-A \right) + {\rm erfc}(A) \;,
\eea
which holds for arbitrary $A$ and $B$. Setting $A = \frac{m_2 - \rho\,m_1}{\sqrt{1-\rho^2}}$ and $B = 2 m_1 \sqrt{1-\rho^2}/\rho$ in this identity (\ref{identity}), we get
\bea \label{IN_large3}
N \,I_N \approx \frac{1}{2}\,e^{-z_1} \left[ e^{\frac{1-\rho^2}{\rho^2} m_1^2 - 2\frac{m_1}{\rho}(m_2 - \rho m_1)}\,{\rm erfc}\left( \frac{m_1}{\rho\sqrt{1-\rho^2}} - \frac{m_2}{\sqrt{1-\rho^2}}\right) + {\rm erfc}\left(\frac{m_2 - \rho\, m_1}{\sqrt{1 - \rho^2}} \right)\right] \;.
\eea
We will now analyse this expression in the large $N$ limit where both $m_1$ and $m_2$ are large and are given by Eq. (\ref{mk_zk}), i.e., we substitute 
\bea \label{m1_m2}
m_1 = a_N + \frac{1}{2\,a_N} z_1 \quad {\rm and} \quad m_2 = a_N + \frac{1}{2\,a_N} z_2 \;,
\eea
where we used $b_N = 1/(2 a_N)$ from (\ref{aNbN}). Here, in the large $N$ limit, $a_N$ is large and is given by (\ref{aN_largeN}), while $z_1$ and $z_2$ are fixed and both are of $O(1)$ for large $N$. We substitute (\ref{m1_m2}) in (\ref{IN_large3}) and expand for large $a_N$ using Mathematica. This gives
\bea \label{IN_large4}
N\,I_N \approx A_N(\rho)\, e^{-\frac{1}{1+\rho}(z_1+z_2)} \quad {\rm where} \quad A_N(\rho) = \frac{1}{2 \sqrt{\pi}} \frac{(1+\rho)^{3/2}}{\sqrt{1-\rho}} \frac{1}{a_N}\, e^{- \frac{1-\rho}{1+\rho}\,a_N^2} \;. 
\eea
Using the second relation in (\ref{aNbN}), the expression for $A_N(\rho)$ simplifies, leading to
\bea \label{AN_rho2}
 A_N(\rho) \approx \frac{1}{2 \sqrt{\pi}} \frac{(1+\rho)^{3/2}}{\sqrt{1-\rho}} \frac{1}{(a_N)^{\frac{2\rho}{1+\rho}}} \left(\frac{N}{2\sqrt{\pi}} \right)^{-\frac{1-\rho}{1+\rho}} \;.
\eea
Finally, using (\ref{aN_largeN}), one gets to leading order for large $N$
\bea \label{AN_rho3}
 A_N(\rho) \approx \frac{1}{2 \sqrt{\pi}} \frac{(1+\rho)^{3/2}}{\sqrt{1-\rho}} \frac{1}{\left(\ln \tilde N\right)^{\frac{\rho}{1+\rho}}} \, {\tilde N}^{-\gamma} \quad {\rm with} \quad \gamma = \frac{1-\rho}{1+\rho} \;, 
\eea
where we recall that $\tilde N = N/(2 \sqrt{\pi})$. Thus Eqs. (\ref{IN_large4}) and (\ref{AN_rho3}) provide the asymptotic large $N$ analysis of $I_N$ in Eq.~(\ref{def_I}), for fixed $z_1$ and $z_2$.

\section{Scaled correlation function in the large $N$ limit} \label{sec:correl}

Armed with the asymptotic behaviour of $I_N$ in Eq. (\ref{IN_large4}), we are now ready to analyse the scaled cumulative distribution ${\cal F}_N(m_1, m_2, \rho)$ in Eq. (\ref{calF_N_SM}) and, from that, extract the two-time correlation function ${\cal C}_N(t_1, t_2)$. We first note that the expression for the joint cumulative 
distribution in Eq. (\ref{calF_N_SM}) can be approximated, to leading order for large $N$, as 
\bea \label{approx_FN}
{\cal F}_N(m_1, m_2, \rho) \approx \exp\left[-\frac{N}{2} {\rm erfc}(m_1) -\frac{N}{2} {\rm erfc}(m_2) + N I_N \right] \;.
\eea
Using the asymptotic behavior of ${\rm erfc}(y)$ in Eq. (\ref{large_erfc}) and the relation in Eq. (\ref{rel_zk2}), we get
\bea \label{approx_FN.2}
{\cal F}_N(m_1, m_2, \rho) \approx \exp\left[-e^{-z_1}-e^{-z_2} + N\,I_N \right] \;.
\eea
Substituting now the behavior of $N I_N$ from Eq. (\ref{IN_large4}), we get
\bea \label{asympt_corr1}
{\cal F}_N(m_1, m_2, \rho) \approx \exp\left[-e^{-z_1}-e^{-z_2} + A_N(\rho)e^{-(z_1+z_2)/(1+\rho)}\right] \;.
\eea
Note that we have also used the relation in Eq. (\ref{m1_m2}) for $m_1$ and $m_2$ in terms of $z_1$ and $z_2$.
 
We now assume that $\rho$ is fixed and take the limit $N \to \infty$. In this limit, using the expression of $A_N(\rho)$ from (\ref{AN_rho3}), we see that the second term in the exponential in Eq. (\ref{asympt_corr1}) decays as $N^{-\gamma}$ with $\gamma < 1$, since $\rho>0$. Using  the fact that $A_N(\rho) \ll 1$ for large $N$, we can now expand the exponential in powers of $A_N(\rho)$. 
Keeping only the leading term, we get
\bea \label{asympt_corr2}
{\cal F}_N(m_1, m_2, \rho) \approx \tilde F_N(z_1, z_2, \rho) = e^{-e^{-z_1}-e^{-z_2}}\left[1+ A_N(\rho) e^{-\frac{1}{1+\rho}(z_1+z_2)}  \right] \;.
\eea

From this joint cumulative distribution of the scaled variables, we can express their joint PDF by taking derivatives with respect to $z_1$ and $z_2$, giving 
\begin{equation} \label{joint_PDFtilde}
\tilde P_N(z_1, z_2, \rho) = \frac{\partial^2 \, \tilde F_N(z_1, z_2, \rho)}{\partial z_1 \partial z_2} = e^{-z_1-z_2-e^{-z_1}-e^{-z_2}} + A_N(\rho) \left(e^{-z_1} - \frac{1}{1+\rho} \right) \left(e^{-z_2} - \frac{1}{1+\rho} \right)\, e^{-\frac{z_1+z_2}{1+\rho}-e^{-z_1}-e^{-z_2}} \;.
\end{equation}
Note that this joint PDF is a completely symmetric function of $z_1$ and $z_2$. The connected two-point correlation of the scaled variables $Z_1$ and $Z_2$ is then given by
\bea \label{scaled_correl}
\langle Z_1\, Z_2\rangle - \langle Z_1\rangle\langle Z_2\rangle = \int_{-\infty}^\infty dz_1 \,\int_{-\infty}^\infty dz_2\, z_1\, z_2\,  \tilde P_N(z_1, z_2, \rho) - \left[ \int_{-\infty}^\infty dz_1 \,\int_{-\infty}^\infty dz_2\, z_1\,  \tilde P_N(z_1, z_2, \rho) \right]^2 \;.
\eea
In the next section, we show that, using $\tilde P_N(z_1, z_2, \rho)$ from Eq. (\ref{joint_PDFtilde}), the integrals over $z_1$ and $z_2$ in (\ref{scaled_correl})
can be performed explicitly, leading to
\bea \label{correl_expl}
\langle Z_1\, Z_2\rangle - \langle Z_1\rangle\langle Z_2\rangle \approx A_N(\rho) \, \Gamma^2\left(\frac{1}{1+\rho}\right) \;.
\eea
Going back to the scaled maxima $\tilde M_1$ and $\tilde M_2$ 
\bea \label{scaled_Max}
\tilde M_1 = a_N + \frac{1}{2 a_N} Z_1 \quad {\rm and} \quad \tilde M_2 = a_N + \frac{1}{2 a_N} Z_2 \;,
\eea
their connected correlator can be expressed as
\bea \label{correl_expl2}
\langle \tilde M_1 \tilde M_2\rangle - \langle \tilde M_1\rangle \,\langle \tilde M_2\rangle = \frac{1}{4\,a_N^2} \left[ \langle Z_1\, Z_2\rangle - \langle Z_1\rangle\langle Z_2\rangle\right] \;.
\eea
Thus, using the result in (\ref{correl_expl}), we get for large $N$
\bea \label{correl_expl3}
\langle \tilde M_1 \tilde M_2\rangle - \langle \tilde M_1\rangle \,\langle \tilde M_2\rangle \approx \frac{A_N(\rho)}{4 a_N^2} \Gamma^{2}\left(\frac{1}{1+\rho}\right) \;. 
\eea
Finally, using the explicit expression of $A_N(\rho)$ from (\ref{AN_rho3}) and of $a_N$ from (\ref{aN_largeN}), we get our final result for the leading large $N$ behavior of the connected correlator of the scaled maximum
\bea \label{correl_final}
{\cal C}_N(t_1, t_2) = \langle \tilde M_1 \tilde M_2\rangle - \langle \tilde M_1\rangle \,\langle \tilde M_2\rangle \approx \frac{B(\rho)}{\left( \ln \tilde N\right)^{\frac{1+2\rho}{1+\rho}}}\, \tilde N^{-\gamma}  \;,
\eea
where the decorrelation exponent $\gamma$ and the amplitude $B(\rho)$ are given explicitly by 
\begin{equation} \label{gamma_rho}
\gamma = \frac{1-\rho}{1+\rho}  \quad, \quad B(\rho) = \frac{1}{8 \sqrt{\pi}} \frac{(1+\rho)^{3/2}}{\sqrt{1-\rho}}\, \Gamma^2\left( \frac{1}{1+\rho}\right) \;.
\end{equation}
This completes the derivation of our main result stated in Eqs. (8) and (9) of the main text.

\section{Computation of the integrals in (\ref{scaled_correl})}

The joint PDF ${\tilde P_N}(z_1, z_2, \rho)$ in Eq. (\ref{joint_PDFtilde}) is of the form
\bea\label{form}
{\tilde P_N}(z_1, z_2, \rho) = f(z_1) f(z_2)+ A_N(\rho) g(z_1) g(z_2) \;,
\eea
where 
\bea \label{fg}
f(z) = e^{-z - e^{-z}} \quad {\rm and} \quad g(z) = \left(e^{-z} - \frac{1}{1+\rho}\right)\,e^{-\frac{z}{1+\rho}-z} \;.
\eea
Note that the function $f(z)$ is normalized to unity, namely
\bea \label{norm_f}
\int_{-\infty}^\infty f(z)\, dz = 1 \;. 
\eea
It is also trivial to see that the function $g(z)$ defined above satisfies the identity
\bea \label{id_g}
\int_{-\infty}^\infty g(z) \, dz = 0 \;.
\eea
Using these two identities we see that
\bea \label{av_z}
\langle Z_1 \rangle = \int_{-\infty}^\infty z_1\, f(z_1)\, dz_1  \quad, \quad \langle Z_2 \rangle = \int_{-\infty}^\infty z_2\, f(z_2)\, dz_2 \;.
\eea
Similarly, one has
\bea \label{id_correl}
\langle Z_1 Z_2 \rangle =  \left( \int_{-\infty}^\infty z\, f(z)\, dz\right)^2 + A_N(\rho)  \left( \int_{-\infty}^\infty z\, g(z)\, dz\right)^2 \;.
\eea
Therefore the connected correlator is given by
\bea \label{connected_co}
\langle Z_1 Z_2 \rangle - \langle Z_1\rangle\langle Z_2 \rangle = A_N(\rho) \,\left( \int_{-\infty}^\infty z\, g(z)\, dz\right)^2 \;.
\eea
Expressing $g(z) = -\partial_z\left(e^{-z/(1+\rho)-e^{-z}} \right)$, the integral in Eq. (\ref{connected_co}) can be performed by parts leading to
\bea \label{connected_co2}
\langle Z_1 Z_2 \rangle - \langle Z_1\rangle\langle Z_2 \rangle = A_N(\rho)  \left[\int_{-\infty}^\infty e^{-\frac{z}{1+\rho}-e^{-z}} \, dz\right]^2 \;.
\eea 
Performing the change of variable $e^{-z} = y$ then gives the result
\bea \label{connected_co3}
\langle Z_1 Z_2 \rangle - \langle Z_1\rangle\langle Z_2 \rangle = A_N(\rho)  \, \Gamma^2\left( \frac{1}{1+\rho}\right) \;,
\eea 
that was used in Eq. (\ref{correl_expl}). 

\section{Connected correlator of the scaled maximum in the limit $\rho \to 0$}

The derivation of the main formula for the connected correlator in Eq. (\ref{correl_final}) assumed that the correlation coefficient $\rho$ is {\it strictly} positive, i.e., $\rho >0$. One natural question is what happens in the limit $\rho \to 0$. From the explicit formula for $A_N(\rho)$ in Eq. (\ref{AN_rho2}), we see that in the limit $\rho \to 0$
\bea \label{AN_small_rho1}
A_N(\rho) \underset{\rho \to 0}{\simeq} \frac{1}{2\sqrt{\pi}} \left(1 + 2\rho + O(\rho^2) \right)\,\frac{e^{(2\rho-1)a_N^2}}{a_N} \;.
\eea
Using now the second relation in Eq. (\ref{aNbN}) for $a_N$, namely, $e^{-a_N^2} \approx 2\sqrt{\pi}\,a_N/N$, one gets
\bea \label{AN_small_rho2}
A_N(\rho) \underset{\rho \to 0}{\simeq} \frac{1}{N} + \frac{2}{N}(a_N^2+1)\,\rho + O(\rho^2) \;. 
\eea
Note that for large $N$, one has $a_N^2 \simeq \ln \tilde N $ from Eq. (\ref{aN_largeN}). This means that, for small $\rho$ and large $N$, the leading behavior of $A_N(\rho)$ in Eq. (\ref{AN_small_rho2}) is then given by
\bea  \label{AN_small_rho3}
A_N(\rho) \underset{\rho \to 0}{\simeq} \frac{2 a_N^2}{N}\,\rho \;.
\eea
Substituting this behavior in Eq. (\ref{connected_co3}) gives the leading behavior as $\rho \to 0$ 
\bea \label{cc_small_rho1}
\langle Z_1 Z_2 \rangle - \langle Z_1 \rangle \langle Z_2 \rangle \approx \frac{2 \rho a_N^2}{N} \;.
\eea
Consequently, the connected correlator of the scaled maximum from Eq. (\ref{correl_expl3}) reads, after cancelling the factor $a_N^2$,
\bea \label{cc_small_rho2}
\langle \tilde M_1 \tilde M_2\rangle - \langle \tilde M_1\rangle \,\langle \tilde M_2\rangle \underset{\rho \to 0}{\to} \frac{\rho}{2\,N} \;.
\eea
This reproduces the result given Eq. (18) in the main text. Using $\tilde M_1 = M_1/\sqrt{2\,C_{11}}$, $\tilde M_2 = M_2/\sqrt{2\,C_{22}}$ and $\rho = C_{12}/\sqrt{C_{11}\,C_{22}}$, the connected correlator for the unscaled maximum, as $\rho \to 0$, is given by
\bea \label{cc_small_rho3}
\langle M_1 M_2\rangle - \langle M_1\rangle \,\langle M_2\rangle \underset{\rho \to 0}{\to} \frac{C_{12}}{N} \;.
\eea
This result has a nice physical interpretation. In general, since the walkers are independent, the corrected correlator of the unscaled maximum can be written
as [see Eq. (5) of the main text]
\bea \label{rel_SN}
\langle M_1 M_2\rangle - \langle M_1\rangle \,\langle M_2\rangle = C_{12}\,S_N(t_1, t_2) \;,
\eea
where $S_N(t_1, t_2)$ is the probability that, given that the maximum at time $t_1$ is achieved by one the walkers, the maximum at time $t_2$ is achieved by the same walker. In general, it is difficult to compute $S_N(t_1,t_2)$ because of the finite correlation present between the two times $t_1$ and $t_2$. However, when $\rho \to 0$, or equivalently $C_{12} \to 0$, this probability $S_N(t_1, t_2) \to 1/N$ since the probability that the leader is the same at time $t_1$ and $t_2$ is simply $1/N$ when there are no correlations. Consequently, substituting $S_N(t_1,t_2) = 1/N$ in Eq. (\ref{rel_SN}) gives, to leading order for small $C_{12}$, the result in Eq. (\ref{cc_small_rho3}).  

\section{Connected correlator of the scaled maximum in the limit $\rho \to 1$}

We now consider the opposite limit $\rho \to 1$, or equivalently $t_1 \to t_2$, where each walker is strongly correlated in time and consequently their global maximum at two different times is also strongly correlated. Taking the limit $\rho \to 1$ in Eq. (\ref{IN_large4}), we get
\bea \label{AN_rho1.1}
A_N(\rho)  \underset{\rho \to 1}{\to} \sqrt{\frac{2}{\pi}} \frac{1}{\sqrt{1-\rho}} \; \frac{1}{a_N} \;.
\eea
Substituting this behavior in Eq. (\ref{correl_expl3}), we find that the connected scaled correlator, behaves, to leading order as $\rho \to 1$ 
\bea \label{rhoto1.1}
{\cal C}_N(t_1, t_2) = \langle \tilde M_1 \tilde M_2\rangle - \langle \tilde M_1\rangle \,\langle \tilde M_2\rangle \underset{\rho \to 1}{\to} \sqrt{\frac{\pi}{8(1-\rho)}} \,\frac{1}{a_N^3} \;.
\eea
This reproduces the result given in Eq. (19) of the main text, once we use $a_N \approx \sqrt{\ln \tilde N}$. 

On the other hand, when $t_1$ is very close to $t_2$, we would expect that the connected correlator should converge to the variance of the scaled maximum as given in Eq. (\ref{mean_var_MN}), i.e., 
\bea \label{rhoto1.2}
{\cal C}_N(t_1, t_2) \underset{\rho \to 1}{\to}  \frac{\pi^2}{24 \,a_N^2} \;.
\eea
Comparing these two behaviors (\ref{rhoto1.1}) and (\ref{rhoto1.2}), we see that, as $\rho \to 1$, the scaled correlator crosses over from the behavior in Eq. (\ref{rhoto1.1}) to the one in Eq. (\ref{rhoto1.2}) when 
\bea \label{rhoto1.3}
\rho = 1 - \frac{\nu}{a_N^2} \;,
\eea
where $\nu = O(1)$ as $N \to \infty$. In order to study this crossover, we analyse the large $N$ behavior of the scaled correlator ${\cal C}_N(t_1, t_2) $ by replacing $\rho$ by Eq. (\ref{rhoto1.3}) keeping $\nu$ fixed. We now substitute (\ref{rhoto1.3}) in the expression for $I_N$ in Eq.~(\ref{IN_large3}) and also replace $m_1$ and $m_2$ in terms of their associated scaled variables $z_1$ and $z_2$ as in Eq. (\ref{m1_m2}). Now taking $a_N \to \infty$ limit, while keeping $\nu$ fixed, we see that $N I_N$ approaches a limiting form independent of $N$, namely
\bea \label{IN_rho1}
N I_N \underset{N \to \infty}{\to} G_\nu(z_1, z_2) = \frac{1}{2} \left[e^{-z_1} \, {\rm erfc}\left(\frac{2\nu -z_1 + z_2}{\sqrt{8\,\nu}} \right) + e^{-z_2}\, {\rm erfc}\left(\frac{2\nu+z_1-z_2}{\sqrt{8 \nu}} \right)\right] \;.
\eea
This scaling function $G_\nu(z_1, z_2)$ is clearly symmetric under the exchange of $z_1$ and $z_2$. We now substitute this scaling form in Eq. (\ref{approx_FN.2}) and get the leading behavior of the joint cumulative distribution 
\bea\label{FN_rho1}
{F}_N(m_1, m_2, \rho) \underset{N \to \infty}{\to} \exp\left[-e^{-z_1} - e^{-z_2} + G_\nu(z_1, z_2) \right] \;.
\eea
This limiting joint cumulative distribution is thus independent of $N$ for large $N$ and is parametrised by $\nu$. This scaling limit, when $\rho \to 1 - \nu/\ln N$, 
has been well studied~\cite{kabluchko,BR1977,HR1989,pierre} using different methods. Our result for the limiting two-time joint distribution in Eq. (\ref{FN_rho1}) agrees with these results. However, we would like to emphasize that, away from this limit when $\rho = O(1)$, our main result stated in Eqs. (8) and (9) of the main text has not been found before. In particular, the power law decay of the scaled correlator $\sim N^{-(1-\rho)/(1+\rho)}$ as a function of $N$ for large $N$ and fixed $\rho$ is novel to our knowledge.